\documentclass[11pt,titlepage]{article}

\usepackage{epsfig,graphicx}

\newcommand {\be}{\begin{equation}}
\newcommand {\ee}{\end{equation}}
\newcommand {\bra}{\langle}
\newcommand {\ket}{\rangle}
\newcommand {\HH}{{\hat H}}
\newcommand {\LL}{{\mathcal L}}
\newcommand {\TT}{T^{\sl t}}
\newcommand {\tTT}{{\tilde T}^{\sl t}}

\title{Variational Estimates using a Discrete Variable Representation}

\author{
M. Lombardi, P. Barletta, A. Kievsky \\
Istituto Nazionale di Fisica Nucleare, Via Buonarroti 2, 56100 Pisa, Italy \\
 and \\
Dipartimento di Fisica, Universit\`a di Pisa, 56100 Pisa, Italy}

\begin{document}

\maketitle

\begin{abstract}

 The advantage of using a Discrete Variable Representation (DVR) is that
 the Hamiltonian of two interacting particles can be constructed
 in a very simple form. However the DVR Hamiltonian is approximate
 and, as a consequence, the results cannot be considered
 as variational ones. We will show that the variational character of the results
 can be restored by performing a reduced number of integrals. 
 In practice, for a variational description of the lowest $n$ bound states
 only $n(n+1)/2$ integrals are necessary
 whereas $D(D+1)/2$ integrals are enough for the scattering states ($D$ is the
 dimension of the $S$ matrix). Applications of the method to the study of 
 dimers of He, Ne and Ar, for both bound and scattering states, are
 presented.

\end{abstract}

\pagebreak

\section{Introduction}

The solution of a quantum mechanical problem can be obtained using a variational
principle. It provides second order estimates of binding energies 
and scattering matrices as well as the corresponding wave functions.
For bound states the Rayleigh-Ritz
variational principle can be used whereas for scattering states
the Kohn variational principle (KVP), among others, can be employed. 
A common procedure is to combine the variational principle with the expansion
of the wave function over a complete basis. Following Ref.\cite{lhl85}
we will refer to this method as the variational basis representation (VBR). 
The accuracy of such a method is strictly connected to the size of the basis 
set employed. In practical
applications the basis set, which in many cases is infinite, is truncated.

For the specific case of bound state calculations the implementation
of the VBR method leads to the solution of a generalized eigenvalue
problem. The corresponding eigenvalues and eigenvectors represent upper
bounds to the exact energy levels of the Hamiltonian and first order
approximations to the associated wave functions. If $N$ is the
dimension of the truncated basis the Hylleras-Undheim theorem guarantees
that
\begin{equation}
 E_{\lambda+1}(N) \ge E_\lambda(N)\ge E_\lambda(N+1)  \;\;\; ,
\label{feq}
\end{equation}
and 
\begin{equation}
 \lim_{N\rightarrow\infty}E_\lambda(N)=\epsilon_\lambda  \;\;\; ,
\end{equation}
where $\epsilon_\lambda$ represents the exact eigenvalue of level $\lambda$.
As there is no approximation involved in the calculation of the
Hamiltonian matrix elements,
the accuracy of the VBR is directly related to the completeness of the basis
employed. Accordingly, by increasing the dimension of the
basis it is possible to obtain solutions extremely close to the exact ones.

For scattering states the situation is slightly different. The wave 
function describing a scattering state is not an $L^2$ function, however 
its form outside the region where the collision
takes place is in general known. Therefore
the configuration space can be divided into two regions: the
asymptotic region in which the particles are either free or
interacting through a long range (coulombic) potential, and the
internal region. In the asymptotic region the scattering wave
function can be described in terms of Bessel or Coulomb functions. In the
internal region the scattering state can be expanded over a complete
$L^2$-basis. 
The linear coefficients of the expansion and the $S$ matrix elements 
giving the
relative weights between the ingoing and outgoing asymptotic solutions 
can be obtained from the KVP.
The implementation of the Kohn variational principle using a VBR leads to a two-step
procedure. The first step consists in obtaining a first order estimate
of the scattering matrix $S$ and of the wave function by solving a linear
non-homogeneous system of equations of dimension $N+D$, with $D$ the
dimension of the $S$ matrix. A second order estimate of the $S$ matrix is
obtained by inserting the first order solution in the functional:
\begin{equation}
 [S_{\alpha\beta}]
= S_{\alpha\beta}+i<\Psi^-_\beta|H-E|\Psi^+_\alpha>\;\;\; ,
\label{sfun}
\end{equation}
with $H$ and $E$ the Hamiltonian and the energy of the system respectively.
$S_{\alpha\beta}$ is the first order estimate of the corresponding 
$S$ matrix element and $\Psi^+_\alpha$ ($\Psi^-_\beta$) 
is the outgoing (ingoing) solution corresponding to channel $\alpha$ ($\beta$).
The convergence of the second order estimate of the $S$ matrix elements 
can be studied for
increasing values of $N$. Although in this case there is no guarantee
of a monotonous convergence, in contrast with the case for bound states, the 
properties of the KVP have been extensively studied in past years~\cite{kvp}.

The key hypothesis in the variational theorems when implemented with the VBR
method is that all the required matrix elements need to be
calculated exactly, or at least to a very high accuracy. This means
that the computational cost of the calculation increases as $N^2$. It
is quite common to choose orthogonal polynomials with appropriate weight
functions as basis sets, so that both the norm and the kinetic energy matrix 
elements are analytical, and the only numerical integrations required are those of
the potential matrix elements.

A different, but related technique, is the
Discrete Variable Representation (DVR) which is widely used in the description
of molecular systems. Examples are the calculations of vibrational spectra,
scattering problems or photo-dissociation processes~\cite{light1,hesse99}.
The DVR method is described in detail in a recent review \cite{DVR}.
The basic property
of the DVR is that local operators depending on the inter-particle
distance (which include the potential) are diagonal in the DVR basis. However, 
in the case of the potential matrix the diagonal form is a result of 
an approximate treatment of the integrals defining the corresponding matrix elements.
For example, in the case of an $N$ dimensional basis of orthogonal polynomials
multiplied by appropriate weight functions,
the approximation consists in replacing the integrals by the related
$N$ point Gaussian quadrature formula. In the limit of $N$ going to infinity,
both the VBR and the DVR approximations for the energy levels converge to
the exact solutions. The advantage of the DVR with respect to the VBR
is that there is no need to compute $N^2$ integrals, and the price to
pay is the loss of the variational character of the calculated eigenvalues and
eigenvectors.

In the present paper we would like to investigate the possibility of producing
variational estimates using the DVR method. For bound states we proceed as
follows. The solution of an $N$ dimensional DVR problem yields 
$N$ eigenvalues $E^{\rm {DVR}}_\lambda$ and the corresponding $N$ eigenvectors
$\psi^{\rm {DVR}}_\lambda$. The lowest eigenvalue $E^{\rm {DVR}}_0$ is
an approximation to the exact ground state $\epsilon_0$. Due to the absence
of the variational character, $E^{\rm {DVR}}_0$ could be either greater
or smaller than $\epsilon_0$ and this ambiguity remains even for large
values of $N$. However the quantity
\be
{\mathcal E}_0=\bra\psi^{\rm{DVR}}_0|H| \psi^{\rm{DVR}}_0 \ket
\ee
is an upper bound to $\epsilon_0$ provided that the above integral
is calculated accurately. To this aim, the $\psi^{\rm {DVR}}_0$ wave
function, which is known in the DVR basis, that is as amplitudes at the DVR
points, has to be transformed to the $r$-space.
As a result, only one numerical integral needs to be evaluated in order
to obtain an upper bound to the ground state energy.
As we will see this procedure can be extended to
produce upper bounds to the lowest $n$ levels belonging to a band
of the exact Hamiltonian. In this case a minimum of
$n(n+1)/2$ integrals have to be computed. In general the number $N$
of basis functions required to describe correctly the first $n$
states of an spectral band satisfies $N >> n$, so that the number of integrals to be
performed in order to produce the variational bounds is much smaller
than those needed with the VBR. In cases where it is possible
to obtain good solutions with $N\approx n$, the
alternatives to the VBR method, such as the one here presented, do not have
any particular advantage since the basis functions used to expand
the wave functions are already close to the exact solutions.

For scattering states a similar procedure can be followed. If $S^{\rm{DVR}}$
is the first order estimate of the $S$ matrix element calculated from
the KVP using the DVR technique,
and $\psi^{\rm {DVR}}$ is the
corresponding scattering state, we can use eq. (\ref{sfun}) to calculate
the second order estimate of the $S$ matrix element. In order to produce a
true variational estimate, the integral of  eq. (\ref{sfun}) must be
calculated accurately. This can be achieved again by transforming the DVR
solution to the $r$-space. In this manner we have obtained
a variational estimate of the $S$-matrix elements computing
$D(D+1)/2$ numerical integrals, where $D$ is the dimension of the $S$-matrix.
In most cases $D$ is much smaller than the dimension of the basis $N$.

In order to illustrate the application of the above formalism,
we have studied dimers formed by two equal
rare gas atoms. The spectrum of small clusters of He, Ne, Ar, Kr and Xe
has been the subject of recent investigations (see Ref.~\cite{noble} and references 
therein). In particular, the study of the Helium dimer has proved very
challenging, due to the extraordinary features of the He-He interaction
that make the He$_2$ molecule very elusive~\cite{helium}.
Difficulties in the theoretical study of rare gas clusters emerge from the 
characteristics of the atom-atom interaction, whose attractive part is of 
the van der Waals type and whose short range part consists of a hard repulsive core, 
causing those systems to be strongly correlated.
As a result, the rotational-vibrational spectrum of these dimers differs
substantially from the spectrum of typical covalent or polar molecules.
The helium dimer has only one vibrational state, no rotational spectrum
and its binding energy is about seven orders of magnitude smaller than that of
traditional molecules. Some potential models predict
three and nine vibrational states, respectively, for the Ne and Ar dimers
however only some of those states
have been observed. When the VBR is used to describe the bound states of
these dimers a large number of basis functions is required to account for
the strong atom-atom correlation. On the other hand the limited number
of bound states makes the van der Waals dimers well suited for
applications of the method outlined here. We have studied
the dimers He$_2$, Ne$_2$ and Ar$_2$ with state-of-the-art atom-atom
potentials.
In Section II we outline the method as applied to bound states.
The convergence patterns of the energies of different states,
as well as of other observables, are shown for comparison.
In Section III we present results for the scattering lengths and the low
energy phase-shifts. Section IV is devoted to the conclusions.

\section{Two particles bound states using VBR and DVR}

The center of mass Hamiltonian operator for two identical particles is
written as
\begin{equation}
   H=-{\hbar^2\over M} \nabla^2 + V(r) \;\;\; ,
\label{sch}
\end{equation} 
where $M$ is the atomic mass and $V(r)$ represents the interaction.
In order to implement the VBR method we introduce the orthonormal basis 
$\Phi_{klm}({\bf r})=\phi_{lk}(r)Y_{lm}(\hat r)$. The radial basis, which in
the following we assume to belong to a family of orthogonal polynomials,  
satisfies
\be
\int_0^\infty \phi_{lk}(r)\phi_{lk'}(r) r^2 dr= \delta_{kk'} \;\;\; .
\ee
For the sake of clarity, and without loss of generality, we can limit the
discussion to $l=0$ states and call the corresponding radial basis elements
$\phi_k$ ($k=0,1,\dots$). The radial part of the wave function
corresponding to the level $\lambda$ 
is expanded in terms of the first $N$ basis elements as
\begin{equation}
   \psi_\lambda (r)=\sum_{k=0}^{N-1} A^\lambda_k\phi_k(r) \;\;\; .
\label{rvbr}
\end{equation} 
The linear coefficients $A^\lambda_k$ and the upper bounds $E_\lambda$ to the
energy levels are obtained from the eigenvalue
problem resulting from the Rayleigh-Ritz principle:
\begin{equation}
   \sum_{k'=0}^{N-1} (H_{kk'}-E_\lambda\delta_{kk'}) A^\lambda_{k'}=0
\end{equation} 
Both the eigenvalues $E_\lambda$ and the corresponding eigenvectors $A^\lambda_{k'}$ 
depend on the size $N$ of the variational problem solved as is explicitly
evident, for instance, from eq. (\ref{feq}). Therefore, a more complete notation
 would be $E_\lambda (N)$ and $A^\lambda_{k'}(N)$.
However, in the following we will drop the 
index $N$ to simplify the notation.

The Hamiltonian matrix elements $H_{kk'}$ are the sum of the kinetic
and potential energy terms
\begin{eqnarray}
   K_{kk'}&=&-{\hbar^2\over M}\int_0^\infty \phi_k(r)
            \left({d^2\over dr^2}+\frac{2}{r}{d\over dr}\right)\phi_{k'}(r)\; r^2 dr 
   \;\;\; ,
\label{knn} \\
   V_{kk'}&=&\int_0^\infty \phi_k(r)V(r)\phi_{k'}(r)\; r^2 dr
   \;\;\; .
\end{eqnarray} 

With an appropriate choice of the basis set,
the integrals corresponding to the kinetic energy elements $K_{kk'}$
can be obtained analytically whereas, in general,
the integrals corresponding to the potential energy elements $V_{kk'}$
are calculated numerically. As mentioned in the
introduction, the accuracy of the numerical integration must be
high enough to prevent error propagation in the computation
of the eigenvalues of $H$. This meets the hypotheses
of the variational theorem and assures convergence from above to the
exact eigenvalues as the dimension $N$ of the matrix increases.
In the present work the relative accuracy in the computation of the
potential energy elements is better than $10^{-7}$. 

In an alternative approach, the DVR offers a computationally more efficient method
for obtaining good estimates of the eigenvalues and eigenvectors of $H$.
Here we briefly introduce the DVR for a basis
of $N$ orthogonal polynomials times the appropriate weight functions~\cite{lhl85}. 
In this case a one-to-one correspondence exists between the basis
 representation and the representation in $N$ Gaussian
quadrature points, however the method is not limited to this 
case~\cite{lhl85,schn97}.
The DVR corresponding to the set of $N$ basis
functions $\phi_k$ can be obtained from the following unitary matrix
\be
T_{k\alpha}=\phi_k(x_\alpha)\sqrt{\omega_\alpha} \;\;\ ,
\label{tmm}
\ee
where $\{x_\alpha\}$, $\{\omega_\alpha\}$ ($\alpha=0,\dots,N-1$)
are the Gaussian points and weights corresponding to the quadrature formula
\be
\int_0^\infty dr r^2\phi_k(r) V(r)\phi_{k'}(r)
     \approx \sum_{\alpha=0}^{N-1} \omega_\alpha
     \phi_k(x_\alpha)V(x_\alpha)\phi_{k'}(x_\alpha) \;\;\ .
\label{vfbr}
\ee
The Gaussian points $\{x_\alpha\}$ are the DVR points and they can be obtained as
the eigenvalues of the $N\times N$ matrix representation of the coordinate
operator $r$
\be
    r_{kk'}=\int_0^\infty dr r^2  \phi_k(r)\, r\, \phi_{k'}(r) \;\;\; .
\ee
The corresponding eigenvectors form the $T$ matrix defined in eq. (\ref{tmm}).

The right hand side of eq. (\ref{vfbr}) defines the finite basis representation
(FBR) approximation to $V_{kk'}$ and is called
$V^{\rm{FBR}}_{kk'}$. It corresponds to the computation of the potential matrix
elements by quadratures. The potential energy operator in the DVR is defined as
\be
V^{\rm{DVR}}_{\alpha\beta}=(\TT V^{\rm{FBR}}T)_{\alpha\beta}=
V(x_\alpha)\delta_{\alpha\beta} \ .
\ee
The DVR kinetic energy may be expressed as
$ K^{\rm{DVR}}=\TT K T $, where $K$ is the matrix whose
elements $K_{kk'}$ are given by eq. (\ref{knn}).
Therefore we have introduced two isomorphic representations of $H$ that
have a very simple form. In $H^{\rm{FBR}}=K+V^{\rm{FBR}}$,
 the potential energy 
matrix has been calculated using an $N$ point quadrature formula, whereas
in $H^{\rm{DVR}}=K^{\rm{DVR}}+V^{\rm{DVR}}$ the potential energy matrix
 is diagonal. In both cases
the kinetic energy matrix can be obtained analytically~\cite{szalay93}.
The two representations are related
by the unitary transformation $T$, so they have the same set of eigenvalues
$\{E^{\rm{FBR}}_\lambda\}=\{E^{\rm{DVR}}_\lambda\}$.
If $\psi^{\rm{FBR}}_\lambda$ is the eigenvector of $H^{\rm{FBR}}$ with eigenvalue
$E^{\rm{FBR}}_\lambda$, then
\be
\psi^{\rm{FBR}}_\lambda(r)=\sum_{k=0}^{N-1} B^\lambda_k\phi_k(r) 
\label{rfbr}
\ee
corresponds to the eigenvector of $H^{\rm{DVR}}$ in $r$-space,
\be
\psi^{\rm{FBR}}_\lambda(r)\equiv
\psi^{\rm{DVR}}_\lambda(r)=\sum_{\alpha=0}^{N-1} C^\lambda_\alpha
\phi^{\rm{DVR}}_\alpha(r)  \;\;\; .
\label{rdvr}
\ee
The coefficients in the above two equations are related to each other by the
expression
\be
B_k^\lambda=\sum_{\alpha=0}^{N-1} T_{k\alpha} C^\lambda_\alpha \;\;\; ,
\ee
moreover the DVR basis functions $\phi^{\rm{DVR}}_\alpha(r)$ in $r$-space are
\be
\phi^{\rm{DVR}}_\alpha(r)=\sum_{k=0}^{N-1} T_{k\alpha}\phi_k(r) \;\;\
(\alpha=0,\dots,N-1)\
 .
\ee
The $r$-space representation of the DVR basis element
 $\phi^{\rm{DVR}}_\alpha(r)$ is a polynomial of degree $N-1$ with
zeros at the DVR points $x_\beta$, with $\beta \ne \alpha$ \cite{szalay93}.

It is important to notice that the linear coefficients $B^\lambda_k$ 
in eq. (\ref{rfbr}) differ from the VBR coefficients $A^\lambda_k$  of 
eq. (\ref{rvbr}) due to the quadrature formula used to calculate
the potential energy matrix. As a consequence, the eigenvalues obtained
cannot be considered as upper bounds. In fact, we will see that
in several cases the energies $E^{\rm{DVR}}_\lambda$ oscillate around the
convergence value $\epsilon_\lambda$. 
In order to obtain variational estimates, we apply the
variational principle
to the set $n$ of DVR eigenvectors $\psi^{\rm{DVR}}_i(r)$, $i=1,...,n$,
which approximate the $n$ lowest levels of an spectral  band of the exact Hamiltonian. 
In this paper we will discuss the case $l=0$, however the method given below can
be applied just as well to other $l$--bands. 
We can build the $n\times n$ matrix ${\mathcal H}$ as
\be
{\mathcal H}_{ij}=\bra\psi^{\rm{DVR}}_i|{K+V}|
\psi^{\rm{DVR}}_j \ket  \;\;\; ,
 \label{varham}
\ee
computing the matrix elements of ${V}$ numerically using the representation
of $\psi^{\rm{DVR}}$ in $r$-space. Thus,
the computational effort of generating the matrix
elements ${\mathcal H}_{ij}$ is proportional to $n(n+1)/2$. 
Consequently, the eigenvalues ${\mathcal E}_{\lambda}$
of ${\mathcal H}_{ij}$
represent real upper bounds to the eigenvalues $\epsilon_{\lambda}$.
Moreover, as a consequence of the variational principle, the following relation
 holds:
\be
\epsilon_\lambda \le E_\lambda (N) \le {\mathcal E}_\lambda (n) \,\,\, .
\label{varine}
\ee
In fact, whereas 
$E_\lambda (N)$ is the VBR eigenvalue calculated using $N$ basis functions,
${\mathcal E}_\lambda (n)$ is the eigenvalue calculated using $n$
specific combinations of the $N$ basis functions and therefore it is obtained
in a reduced Hilbert space.
This procedure of generating variational bounds allows for a noticeable gain in
computational time with respect to the VBR when the condition $n<<N$ is
satisfied. In principle this condition could seem to be a restriction, but in 
many cases it is valid. For weakly bound molecules, as dimers of rare
gases, the number of bound states is not very high. On the other hand,
for systems having a large number of bound states, 
the number of levels which are approximated well using $N$ basis functions 
is generally smaller than $N$. 

In order to illustrate some applications of the above formalism we will calculate
the bound states
of dimers of He, Ne and Ar. 
For the He-He and Ne-Ne interactions we use the LM2M2 potential 
and the HFD-B potential, respectively, both proposed by 
Aziz and Slaman~\cite{lm2m2,hfd-b}. For the Ar-Ar system we use the HFD-C
 potential proposed by Aziz~\cite{hfd-c}.
The values $\hbar^2/ M=43.281307$, $8.584089$ and $4.336093$ 
${\rm K}\, {\rm a.u.}^2$ have 
been used for the He, Ne and Ar systems, respectively.  
With our choice of the inter-atomic potentials, and of the atomic mass values, 
the number of vibrational states that result is one for He$_2$, 
three for Ne$_2$, and nine for Ar$_2$. The corresponding energy values are
given in Table I.

For the radial basis functions we have used the following orthonormal basis
\begin{equation}
   \phi_k(r)=\sqrt{\beta^3\over (k+1)(k+2)}\;\;{\cal L}^{(2)}_k(z)\;\;
   {\rm e}^{-z/2}
\end{equation} 
where ${\cal L}_k^{(2)}$ is generalized Laguerre polynomial depending
on $z=\beta r$ with $\beta$ a nonlinear parameter, which can be varied
to improve the convergence patterns.

The kinetic terms of eq. (\ref{knn}) have been evaluated using the following
 analytical form:
\be
   K_{kk'}=-\frac{\beta^2 \hbar^2}{M} \left[ \frac{1}{4}(k+1)(k+2)\delta_{kk'} -\frac{1}{3}k^3 - \frac{3}{2} k^2 - \frac{13}{6}k -1  \right] \ \ \ (k\le k'). 
\label{kan}
\ee

The non-linear parameter $\beta$ was chosen to be 1 a.u.$^{-1}$, 5 a.u.$^{-1}$ and
 10 a.u.$^{-1}$ for He$_2$, Ne$_2$ and Ar$_2$, respectively.  

In figure 1 the convergence of the ground
state energy of He$_2$ is shown as a function of $N$ for the three methods,
VBR, DVR and the mean value calculated using the DVR wave function which in
the following has been called $\bra \rm {DVR} \ket$.
It may be observed that the DVR energy $E_0^{\rm{DVR}}$ (dashed line) oscillates
around the VBR energy $E_0$ (solid line) even at very high values of $N$. The solution
becomes stable for $N>500$ revealing the particular structure of the He$_2$.
On the other hand the $\bra \rm {DVR} \ket$ energy ${\mathcal E}_0$ (dotted line)
has a much more stable pattern of convergence and already at $N=200$ its value
coincides to five digits with the VBR result. In fact, in the figure, the VBR and 
$\bra \rm {DVR} \ket$ curves are almost indistinguishable.
The $\bra \rm {DVR} \ket$ calculations can be
used to estimate the quality of the DVR wave function. In Table II the total
energy, the potential energy and the square root radius are given as
functions of $N$ for the three methods. As already mentioned, we observe a
better stability in the VBR and $\bra \rm {DVR} \ket$ energies. Conversely
the DVR and the $\bra \rm {DVR} \ket$ potential energy and square root radius
show small differences with respect to the VBR values.

In figure 2 the patterns of convergence of the energy are shown for four selected 
vibrational states of Ar$_2$ and Ne$_2$, as a function of the number $N$ of DVR points.
For Ne$_2$ the HFD-B potential predicts three bound states, however the highest
excited state is very loosely bound and has been initially excluded
from the analysis. Therefore the
wave functions $\psi^{\rm{DVR}}_\lambda$ ($\lambda=0,1$) have been used
to calculate the matrix elements of the Hamiltonian as defined in
eq. (\ref{varham}). Accordingly an eigenvalue problem of dimension $n=2$ has
been solved. For Ar$_2$ the HFD-C potential predicts nine bound states, but again the
highest excited state has been initially excluded, therefore
an eigenvalue problem of dimension $n=8$ has been solved.
We have obtained a good convergence for all the state analyzed for
both molecules, but for clarity here we limit the figure  to
show the energies corresponding to the ground states of the two molecules, the 
first excited state of Ne$_2$, and the eighth level of Ar$_2$. 
The convergence pattern of these states present some characteristics that 
merit discussion. As expected
the VBR has a very stable convergence in all cases. The DVR energies approach the
VBR result with a very fast convergence yielding oscillations of
decreasing amplitude as $N$ increases. 
Though the DVR results are qualitatively similar
in the four cases presented, and also in all the other states examined, the
$\bra \rm {DVR} \ket$ results for the two ground states have different patterns.
Whereas ${\mathcal E}_0$ for Ne$_2$ (dotted line in panel (a)) presents a marked
oscillatory pattern, ${\mathcal E}_0$ for Ar$_2$ (dotted line in panel (c)) 
presents a convergence extremely close to that of the VBR case. Furthermore, the two 
$\bra \rm {DVR} \ket$ excited states shown in figure 2 as dotted lines in 
panels (b) and (d) also present a marked oscillatory pattern. 
This non monotonic convergence is not in contradiction with the variational 
principle since the relation presented in eq. (\ref{varine}) holds for each
value of $N$. Moreover these oscillations are a consequence of the non--variational 
character of the DVR wave functions since in many cases we found that
the $N+1$ result is worse than the $N$ result.

It is possible to improve the convergence of the $\bra \rm {DVR} \ket$ by increasing
the number of DVR functions included in the variational problem of eq. 
(\ref{varham}). If all the N DVR wave functions are included in eq. 
(\ref{varham}), the VBR and $\bra \rm {DVR} \ket$ methods become isomorphic.
Whereas a minimum of $n$ basis functions is sufficient to produce 
upper bounds to the $n$ lowest states of a band, it is reasonable to expect
that on increasing the number of functions $\psi^{\rm{DVR}}_\lambda$
the convergence pattern will improve. Let us call this number $n'$. In 
figure 3 the convergence of the two energy levels of  Ne$_2$ (panels (a) and (b))
and the eighth level of Ar$_2$ (panel (c)) are given as a function of 
the number $N$ of DVR points, for different sizes $n'$ of the $\bra \rm {DVR} \ket$ 
problem. We consider the cases $n'=4,8,12$ for Ne$_2$
and $n'=12,20,30$ for Ar$_2$, corresponding to the solid, dashed and 
dotted-dashed lines, respectively. In all cases the plot of the VBR results is shown 
as a dotted line. The eighth level of Ar$_2$ represents the most difficult case
for the $\bra \rm {DVR} \ket$ method as it requires $n'$ to be consistently larger
than $n$ in order to have a smooth convergence pattern.
However, we observe that there is a significant 
improvement in the convergence pattern with choices of $n'$ that still 
satisfy the important condition $n' \ll N$. Moreover, as expected from the 
inequality of eq. (\ref{varine}), the ${\mathcal E}_\lambda$ energies are
always greater than the correspondening $E_\lambda$ energies.

\section{scattering states of two particles using VBR and DVR}

In order to produce variational estimates for the scattering matrix
using the DVR method we use the Kohn variational principle in its
general form \cite{kvp}. Here we give a
brief introduction to the method, limiting it to the case of local central
potentials. Without loss of generality, we treat specifically
the case $l=0$ and the case in which the collision proceeds along one
open channel. In this case the scattering matrix is a scalar quantity.
 The formalism can be
easily generalized to treat more channels and the case in which the long
range Coulomb potential is present \cite{kiev97}.

The radial scattering wave function corresponding to a process at
energy $E$ can be written as a sum of two terms:
\be
\psi^+(r)=\psi_c(r)+\psi^+_a(r)  \,\,\, . 
\label{wf1}
\ee
The first term, $\psi_c$ is the internal part and describes the system 
when the two particles are close to each other. It can be expanded in 
terms of $N$ $L^2$ basis functions just as for the bound states
\be \psi_c(r) = \sum_{k=0}^{N-1} A_{k} \phi_k(r) \;\;\ .
\label{expm}
\ee
The second term describes a general asymptotic scattering state of two particles and
is defined as 
\be
\psi^+_a(r)=\Omega_0(r)+{\mathcal L} \Omega_1(r) \,\,\, ,
\ee
where
\begin{eqnarray}
\Omega_0&=&\sqrt{\frac{Mq}{2\hbar^2}}[u_{00}j_0(qr)+u_{01}{\tilde y}_0(qr)] \,\,\ , \\
\Omega_1&=&\sqrt{\frac{Mq}{2\hbar^2}}[u_{10}j_0(qr)+u_{11}{\tilde y}_0(qr)] \,\,\ .
\end{eqnarray}
are asymptotic scattering states,
with $q^2=\frac{M}{\hbar^2}E$. They are given as combinations of the regular
spherical Bessel function $j_0$ and the product of the irregular
Bessel function $y_0$ and a regularizing factor, namely
\be
{\tilde y}_0(qr)=(1-e^{-\gamma r})y_0(qr) \,\,\, .
\ee
The specific form of the regularizing factor is not crucial
provided that the regularization is made in the internal
region and ${\tilde y}_0\rightarrow y_0$ outside the range of the
interaction. The quantity ${\mathcal L}$ expresses the relative weight of
the two scattering asymptotic states $\Omega_0$ and $\Omega_1$.
The coefficients $u_{ij}$ form a matrix that can be chosen in accordance
with the different meanings of the quantity ${\mathcal L}$. For example the
choices
\begin{eqnarray}
{\cal L} & =R     \quad{\rm for}\quad u= \left( \begin{array}{cc}
                                   1 & 0 \\ 0 & 1\end{array}\right) \ , \\
{\cal L} & =S     \quad{\rm for}\quad u= \left( \begin{array}{cc}
                                   i &-1 \\ i & 1\end{array}\right) \ ,
\end{eqnarray}
define the reactance matrix $R$ and the scattering matrix $S$, respectively. 

The generalized KVP states that the functional 
\be 
[{\mathcal L}]= {{\mathcal L}}- {2\over {\rm det}(u)}
\bra \psi^-|{\hat H}|\psi^+ \ket 
\label{kvpa}
\ee
is stationary with respect to variations of the parameters used to
construct the wave function.
We define ${\hat H}$ to be ${\hat H}= (H-E)$, ${\rm det}(u)$ is the determinant
of the matrix $u$ and $\psi^-$ is
the complex conjugate of $\psi^+$. The normalization
of the asymptotic states is defined so as to satisfy
\be
\bra \Omega_0|\HH |\Omega_1\ket-\bra \Omega_1|\HH |\Omega_0\ket= 
\frac{1}{2}{\rm det}(u) \;\;\; .
\label{norm}
\ee

The unknowns in the wave function of eq. (\ref{wf1}) are the N linear
coefficients $A_{k}$ and the quantity ${\mathcal L}$. The variation of the
 Kohn functional with respect to these unknowns
leads to the following $N+1$ inhomogeneous
system of equations 
\begin{eqnarray}
\sum_{k=0}^{N-1} A_{k} \bra \phi_k|\HH |\phi_{k'}\ket -
{\mathcal L} \bra \phi_{k'}|\HH |\Omega_1\ket &=& -\bra \phi_{k'}|\HH |\Omega_0\ket 
\,\ , \,\,\,\ \ \  (k'=0,\dots,N-1) \,\,\, ,
\nonumber \\
\sum_{k=0}^{N-1} A_{k} \bra \phi_k|\HH |\Omega_1\ket -
{\mathcal L} \bra \Omega_1| \HH |\Omega_1\ket &=& {-{\rm det}(u)
+2 \bra \Omega_1| \HH |\Omega_0\ket + 2 \bra \Omega_0| \HH |\Omega_1\ket\over 4} 
\;\ .
\label{lsys}
\end{eqnarray}
The coefficients $A_{k}$ as well as the first order estimate of
the scattering matrix ${\mathcal L}$ are obtained from the solution of 
the above system of equations. The second order estimate $[{\mathcal L}]$ 
is calculated substituting the first order solution in eq. (\ref{kvpa}) and gives
\begin{eqnarray}
[{\mathcal L}]&=& {\mathcal L}-{2\over {\rm det}(u)}
   \Big\{ \bra \Omega_0|V|\Omega_0\ket +
{\mathcal L}^2 \bra \Omega_1|\HH |\Omega_1\ket  +
{\mathcal L} [\bra \Omega_0|\HH |\Omega_1\ket + \bra \Omega_1|\HH |\Omega_0\ket]
 \nonumber \\
&+&2{\mathcal L}\sum_{k=0}^{N-1} A_{k}\bra \phi_k|\HH |\Omega_1\ket
+2\sum_{k=0}^{N-1} A_{k}\bra \phi_k|\HH |\Omega_0\ket \Big\} \;\;\; .
\end{eqnarray}

The convergence properties of the KVP variational principle have been
extensively studied \cite{schwartz,lucchese}. 
Occasionally singularities occur in its solution,
however the complex form (${\mathcal L}\equiv S$) has a much
more stable convergence pattern. Here we use both forms and
check the consistency of the results by means of the relation
\be
S=(1+iR)(1-iR)^{-1}
\label{smatrix}
\ee
which holds only for the exact matrices. Moreover the unitary condition for
the $S$ matrix, $SS^\dagger=I$ is checked as an indication of
the completeness of the basis used for the expansion of the internal part
of the wavefunction.

The DVR method can be used to calculate the first order estimate of
the scattering matrix. For this we write the linear system of eq. (\ref{lsys}) 
in the compact form 
\be
{\tilde H} {\bf A} = {\bf b} \;\;\; ,
\label{csys}
\ee
where ${\tilde H}$ is the $(N+1) \times (N+1)$ matrix defined to be
${\tilde H}_{kk'}=\bra \phi_k|\HH |\phi_{k'}\ket$ ($k,k'=0 \dots N$) and writing
$\phi_N\equiv\Omega_1$. Accordingly, the last element of the
vector of coefficients ${\bf A}$ is $A_{N}\equiv - {\mathcal L}$ and the
elements of the vector ${\bf b}$ are defined to be
\begin{eqnarray}
b_k&=& -\bra \phi_k|\HH |\Omega_0\ket \hspace{1cm} \;\;\; k<N  \,\,\, ,  \\
b_N&=&\frac{-{\rm det}(u)+2 \bra \Omega_1| \HH |\Omega_0\ket + 2 \bra \Omega_0| 
\HH |\Omega_1\ket}{4} \;\;\; .
\end{eqnarray}
The latter can be reduced to $b_N=\bra \Omega_0 | \HH |\Omega_1 \ket$ 
utilizing eq. (\ref{norm}. 

We can now extend the unitary transformation $T$ which acts in the 
$N\times N$ space to the $(N+1)\times (N+1)$ space by defining the following
unitary matrix $ {\tilde T}$
\begin{eqnarray}
&{\tilde T}&_{ij} = T_{ij} \;\;\; \hspace{2cm} i,j=0\dots N-1 \,\,\, , \nonumber \\
&{\tilde T}&_{Nj} =  {\tilde T}_{jN}= 0 \hspace{1.5cm} j=0\dots N-1 \,\,\, , \\
&{\tilde T}&_{NN}  =  1  \,\,\, .   \nonumber
\end{eqnarray}

As in the case of the bound state, we first introduce ${\tilde H}^{\rm{FBR}}$
and ${\bf b}^{\rm{FBR}}$
in which the integrals involving the potential energy operator are
calculated using an $N$ point quadrature formula. 
The integrals involving the asymptotic functions $\Omega_p$ are also
calculated using quadratures. 
The corresponding vector of linear coefficients is ${\bf A}^{\rm{FBR}}$. 
The DVR representation is defined to be
\be
{\tilde H}^{\rm{DVR}}{\bf A}^{\rm{DVR}}={\bf b}^{\rm{DVR}} \,\,\, ,
\label{dvrsys}
\ee
with 
\begin{eqnarray}
{\tilde H}^{\rm{DVR}}&=& \tTT{\tilde H}^{\rm{FBR}}{\tilde T} \,\,\, , \\
{\bf A}^{\rm{DVR}}&=& \tTT {\bf A}^{\rm{FBR}}  \,\,\, ,\\
{\bf b}^{\rm{DVR}}&=& \tTT {\bf b}^{\rm{FBR}}\,\,\, . 
\end{eqnarray}
The structure of the 
symmetric matrix ${\tilde H}^{\rm{DVR}}$ is the following
\begin{eqnarray}
{\tilde H}^{\rm{DVR}}_{ij}&=&K^{\rm{DVR}}_{ij}+V(x_i)\delta_{ij} \;\;\;
i,j=0\dots N-1 \,\,\, ,\\
{\tilde H}^{\rm{DVR}}_{Nj}&=&\sqrt{w_j}\,[\HH\Omega_1]_{x_j} \;\;\;
j=0\dots N-1\,\,\, , \\
{\tilde H}^{\rm{DVR}}_{NN}&=&\sum_\alpha^{N-1}w_\alpha
\Omega_1(x_\alpha)[\HH\Omega_1]_{x_\alpha}  \,\,\, .
\label{dvrh}
\end{eqnarray}
The structure of the inhomogeneous term ${\bf b}^{\rm{DVR}}$ is
\begin{eqnarray}
b^{\rm{DVR}}_j&=&-\sqrt{w_j}\,[\HH\Omega_0]_{x_j} \;\;\;
j=0\dots N-1 \\
b^{\rm{DVR}}_N&=&-\sum_\alpha^{N-1}w_\alpha
\Omega_0(x_\alpha)[\HH\Omega_1]_{x_\alpha}  \;\;\; ,
\label{dvrb}
\end{eqnarray}
where $[\HH\Omega_p]_{x_\alpha}$ indicates the evaluation of the function
$\HH\Omega_p$ at the DVR point $x_\alpha$.
From the above relations we see that only two quadratures have to be
performed in order to construct the DVR system of equations. The
computational effort is limited as compared to the VBR technique.

The solution $\LL^{\rm{DVR}}= -A^{\rm{DVR}}_{N}$ represents the DVR 
first order estimate of the scattering matrix. It coincides with the solution 
$\LL^{\rm{FBR}}=-A^{\rm{FBR}}_{N}$
obtained by solving the FBR system. However both differ from the
first order scattering $\LL$, obtained by solving the system of
eqs. (\ref{lsys}) using the VBR, due to the quadratures introduced to calculate
the matrix elements. 
It is important to notice that
the asymptotic integrals $\bra \Omega_1| \HH |\Omega_1\ket$ and
$\bra \Omega_0| \HH |\Omega_1\ket$,
involved in the construction of the matrix element ${\tilde H}^{\rm{DVR}}_{NN}$ 
and in the inhomogeneous term $b^{\rm{DVR}}_N$ respectively, do not depend on 
the radial basis $\phi_k$ used to expand the scattering wave function.
When the VBR technique is used, these integrals are calculated numerically using
standart techniques.
In the DVR, as it is clear from eqs. (\ref{dvrh}) and (\ref{dvrb}), they
depend on the dimension $N$ of the basis. In some cases, the use of the VBR values in
${\tilde H}^{\rm{DVR}}_{NN}$ and $b^{\rm{DVR}}_N$, instead of those obtained from
the quadrature formula, fails to yield the solution (an example will be given
at the end of Section III-A). Conversely,
the use of the formulas of eqs. (\ref{dvrh}) and (\ref{dvrb}) produces a convergence 
pattern similar to that of the bound states. This means that one has to be
very careful in mixing exact integrals and quadratures when using the DVR.

In order to produce a variational estimate of the scattering matrix we
construct the DVR radial wave function in $r$-space,
\be
\psi^{\rm{DVR}}(r)=\psi^{\rm{DVR}}_c(r)+\Omega_0(r) \,\,\,
+\LL^{\rm{DVR}}\Omega_1(r) \,\,\, , 
\ee
where the internal wave function is
\be
\psi^{\rm{DVR}}_c(r)=\sum_{\alpha=0}^{N-1} A^{\rm{DVR}}_{\alpha}\phi^{\rm{DVR}}
_\alpha(r) \,\,\, .
\label{dvrwfs}
\ee
Using eq. (\ref{dvrwfs}) and eq. (\ref{kvpa}) it is possible to calculate
 the second order  estimate of the scattering matrix:
\be
[\LL]^{\rm{DVR}}= \LL^{\rm{DVR}}-{2\over {\rm det}(u) } 
\bra \psi^{\rm{DVR}}|{\hat H}|\psi^{\rm{DVR}} \ket \;\;\; .
\ee
The integral 
$\Delta = \bra \psi^{\rm{DVR}}|{\hat H}|\psi^{\rm{DVR}} \ket$ 
converges since ${\hat H}|\psi^{\rm{DVR}} \ket
\rightarrow 0$ for $r>r_I$, where $r_I$ is the interaction range. 

We now consider $[\LL]^{\rm{DVR}}$ to be a second order variational estimate
of the scattering matrix. It has been obtained calculating one integral,
namely the integral $\Delta$. For the case in which $D$ channels are open,
$D(D+1)/2$ integrals must be evaluated. Since, in general, $D \ll N$ the
computational effort needed to produce the DVR variational estimate is much
smaller than that of the corresponding VBR. In the following subsections
 we discuss the results for zero and positive energy scattering separately.

\subsection{zero energy case}

For zero energy scattering the asymptotic wave functions
and the KVP assume a particular form. The zero energy wave function is
\be
\psi_0(r)=\psi_c(r)+\psi_a(r)  \,\,\, , 
\label{wf0}
\ee
with
\begin{eqnarray}
\psi_c(r) &= &\sum_{k=0}^{N-1} A_{k} \phi_k(r) \\
\psi_a(r)&=&\Omega_0(r)-a\Omega_1(r) \;\;\; .
\end{eqnarray}

The asymptotic functions are defined to be
$\Omega_0=\sqrt{{M\over 2\hbar^2}}$ and 
$\Omega_1=\sqrt{{M\over 2\hbar^2}}(1-e^{-\gamma r})/r$ 
and $a$ is the scattering length.
The Kohn functional for the scattering length can be obtained from
eq. (\ref{kvpa}) in which ${\mathcal L}\equiv R$
and taking the limit
\be
\lim_{q\rightarrow 0} {{\rm tan}\,\delta\over q}=-a \;\;\;
\ee
where $\delta$ is the phase shift and we have used the definition
$R={\rm tan}\,\delta$. We obtain
\be 
[a]= {a}+ 2 \bra \psi_0|H|\psi_0 \ket \;\;\ .
\label{kvpo}
\ee
The minimization of the above functional with respect to the set of
coefficients $\{A_k\}$ and the scattering length $a$ leads to a
linear system of equations formally equal to that of eq. (\ref{csys})
or, in the case of the DVR, to that of eq. (\ref{dvrsys}).
We solve both systems to calculate the VBR and DVR first order
scattering lengths for the He$_2$, Ne$_2$ and Ar$_2$ systems. As mentioned
before, the second order estimates are obtained by substituting the first
order solution in eq. (\ref{kvpo}) calculating the
integral $\bra \psi_0|H|\psi_0 \ket$ numerically.
Again we use the label $\bra \rm {DVR} \ket $ for the second order estimate
when using the DVR wave function.

In figure 4 (left panel) we show the He-He scattering length (second order) calculated
using VBR (solid line) and $\bra \rm {DVR} \ket$ (dashed line) as a function 
of the number $N$ of basis functions. The converged quantity is
$a=189.518$ a.u.. The attempt to correct the first order DVR result by
calculating the second order estimate using quadratures produces a
 non-convergent
value as is shown in the figure  with the dotted line. In the following we call
this result DVR$_q$. To further analyze
the differences between the first and second order, in the right panel of
figure 4 we show the VBR and DVR first order results. Whereas the first order
VBR result converges as the dimension of the problem increases, the first
order DVR result oscillates around the exact value even for $N>300$.
By comparing the first order DVR result, the DVR$_q$ result and the
$\bra {\rm DVR} \ket $ result, we conclude that the
error introduced by the quadratures in the integral $\bra \psi_0|H|\psi_0 \ket$
is of the same order of magnitude as the integral itself,
therefore it cannot give the correction properly. 
Conversely, using the DVR as a trial wave function the application of 
the Kohn functional naturally produces a better approximation.
Finally, in figure 5 we show the convergence as a function of $N$
of the (second order) scattering lengths
for the Ne-Ne system (a) and Ar-Ar system (b). The converged values are
$a=28.411$ a.u. and $520.22$ a.u., respectively.
The solid, dashed and dotted lines represent the VBR, $\bra {\rm DVR}\ket$
and the DVR$_q$ second order estimates, respectively. 
For Ne the three calculations converge reasonably well. On the contrary,
in the case of Ar the VBR presents a fast convergence, the $\bra {\rm DVR}\ket$
converges after marked oscillations and the DVR$_q$ fails to converge even
for $N\approx 200$. This is a good example of where the corrective term
proportional to $\Delta$ cannot be calculated using quadratures.
The DVR wave functions for the He-He, Ne-Ne and Ar-Ar systems are given in figure 6
corresponding to cases $N=300$, $N=200$ and $N=400$ respectively.
Each function presents a number of nodes equal to the number of bound states
supported. For the Ar-Ar system a basis of big dimension is needed ($N=400$)
in order to produce a wave function orthogonal to the very loosely $E_8$ state.
The values used for the non-linear parameters are $\beta=1$ a.u.$^{-1}$ and
 $\gamma=0.1$ a.u.$^{-1}$ for He, $\beta=5$ a.u.$^{-1}$ and $\gamma=0.1$ 
 a.u.$^{-1}$ for Ne, and $\beta=5$ a.u.$^{-1}$ and $\gamma=0.5$ a.u.$^{-1}$ for Ar.

 Finally we would like to discuss the definition of the elements
 ${\tilde H}^{\rm{DVR}}_{NN}$ and $b^{\rm{DVR}}_N$ given in
 eqs. (\ref{dvrh}) and (\ref{dvrb}). They correspond to the asymptotic integrals 
 $\bra \Omega_1| \HH |\Omega_1\ket$ and $\bra \Omega_0| \HH |\Omega_1\ket$
 obtained using quadratures and, because of that, they vary with N. 
 Therefore at each successive step the DVR problem has been solved using different
 values for those elements. On the other hand, using the VBR, these
 elements has been computed numerically and independently of $N$.
 The differences can be small but significant, for example in some of the
 present calculations the relative difference between both 
 types of integration is about $10^{-5}$, for $N\approx 200$.
 Analyzing the scattering length convergence pattern as a function of $N$
 using both types of integration we observed small differences in the cases of
 He and Ar. However the quadrature formula is to be preferred since it
 provides a first order estimate slightly closer to the final result.
 In the case of Ne the use of the quadrature formula is mandatory, since
 otherwise the first order estimates are far from any reasonable
 value, even for large values of $N$. This may be understood by comparing
 the scattering lengths with the interaction range $r_I$ (estimates
 of $r_I$ are given below in B). For He and Ar the scattering length $a$
 is much larger than the interaction range whereas this is not the case for Ne. 
 In the latter case
 there is a delicate cancellation between the internal and the
 asymptotic part of the wave function in the region $r_I<r<a$. Using DVR
 this cancellation can be properly achieved only when the definitions of
 eqs. (\ref{dvrh}) and (\ref{dvrb}) are used.

\subsection{positive energy case}
Here we study the convergence properties of the $l=0$ phase shift $\delta$ at
very low energies using the DVR.
In this energy region the phase shift and the scattering length are related by
the effective range expansion
\be
    {q\over {\rm tan}\,\delta}=-\frac{1}{a}+\frac{r_0}{2}q^2 \;\;\; ,
\label{efr}
\ee
where $r_0$ is the effective range parameter defined to be
\be
   r_0={2\over a^2}\int_0^\infty [(r-a)^2-\frac{2\hbar^2}{M}(r\psi_0)^2]\, dr 
   \,\,\, ,
\ee
with $\psi_0$ given in eq. (\ref{wf0}). Since
$(r\psi_0)\rightarrow \sqrt{{M\over 2\hbar^2}}(r-a)$ for values of $r>r_I$,
the above integral goes very rapidly to zero outside the interaction region. 
Therefore $r_0$ is itself a measure of the interaction range $r_I$. 
We obtained the values of $13.94$ a.u., $15.65$ a.u. and $59.25$ a.u.
for He, Ne and Ar, respectively.

In Table III we study the convergence of the phase shifts $\delta$
for the He-He and Ne-Ne systems at $50$ mK and for the Ar-Ar system
at $2$ mK. The VBR, DVR$_q$ and $\bra \rm {DVR} \ket$
results correspond to second order estimates obtained replacing the solutions 
of the linear systems of eqs. (\ref{csys},\ref{dvrsys}) in the Kohn functional 
of eq. (\ref{kvpa}). 
Here we do not want to discuss the convergence properties of the
KVP, which can be found in Refs.~\cite{schwartz,lucchese}. However, in order to avoid
possible singular solutions, we have applied the KVP to the two cases,
${\mathcal L}\equiv R$ and ${\mathcal L}\equiv S$, and checked the equivalence of 
the results in accordance with eq. (\ref{smatrix}). In addition we have verified that
on increasing the number of basis functions
the quantity $|SS^\dagger-I|\approx 10^{-8}$ in all the cases considered,
which is of the same order of magnitude as the differences between the results
using ${\mathcal L}\equiv R$ or ${\mathcal L}\equiv S$. From the table we can see
that the DVR$_q$ phase shifts have a slower convergence pattern in all cases. 
For the He-He
system we have seen that at zero energy the DVR$_q$ result is not stable even
for large $N$ values. At $50$ mK the situation is slightly improved though the
pattern of convergence of DVR$_q$ is still not satisfactory. For Ne and
Ar we observe that big values of $N$ are necessary to obtain stable results. 
This is due to the presence of excited states close to threshold which are
generally difficult to describe, as in the case of the $E_8$ state of Ar.

In figure 7 we compare the quantity $q/{\rm tan}\,\delta$ (filled circles)
to the r.h.s of eq. (\ref{efr}) calculated using the zero energy results
(solid line). From this plot it is possible to extract the energy at which
the phase shift starts to deviate from a linear relation. Below this
energy the collision is not very sensitive to the details of the potential
since the dynamics is governed by two parameters, the scattering length
and the effective range. Greater sensitivity to the potential appears
above this energy. For the He system the deviation from linearity starts at
about $E=0.1$ K whereas for Ne it appears at about $E=0.05$ K and for Ar around
$E=0.002$ K.

\section{Conclusions}

 In the present paper we have studied the possibility of using
 DVR wave functions to produce variational estimates to binding
 energies and scattering matrices. The main advantage in using
 the DVR in the description of a quantum mechanical problem
 is the simplicity in constructing the Hamiltonian matrix.
 However the DVR eigenvalues do not represent upper bounds to
 the exact levels. Therefore, 
 in order to obtain estimates to the levels,
 it is necessary to produce a
 convergence pattern in terms of the number $N$ of DVR points
 and analyze the stability of the results as $N$ increases.
 In the case of the He-He system we have seen that the DVR
 eigenvalue oscillates around the exact level even for high 
 values of $N$. On the contrary, the variational estimate obtained
 with the DVR wave function shows a better stability. As a result,
 calculating only one integral namely the mean value of the Hamiltonian with
 the DVR ground state wave function, we have improved the prediction using
 the DVR method. 

 In order to extend the discussion to excited states,
 we have studied the Ne-Ne and Ar-Ar systems. We have shown that we
 need to perform at least $n(n+1)/2$ integrals to produce variational
 estimates of the first $n$ levels. Although the variational estimates
 are always upper bounds, we have noticed that, for some levels, they 
 do not converge monotonically, presenting instead oscillations. 
 In order to improve this behavior we have considered a variational problem
 with an enlarged number ($n'$) of DVR wave functions.
 We have seen that with $n' \approx 2n$ we obtain convergence patterns 
 very close to those obtained with the VBR. In the case in which
 the description includes many excited states with a big spread in energy,
 we have found it necessary to further increase $n'$. Both $n$ and $n'$
 are in general much smaller than $N$, so the possibility of generating
 variational bounds using the DVR is convenient as compared to the VBR,
 from the point of view of the computational effort required.

 For scattering states the Kohn Variational Principle gives a natural context
 in which the DVR wave function may be used to calculate second order variational 
 estimates of the scattering matrix. The case of the He-He system at zero energy
 is a good example. The
 first order scattering length calculated using DVR oscillates around the
 exact value. The second order variational estimate calculated using the DVR
 wave function has a very fast convergence. Conversely, the second order
 when calculated
 using quadratures fails to reproduce the correct value for $N<500$. 
 This is a consequence
 of the particular structure of the He-He interaction. For the Ne-Ne 
 system the differences between the second order calculated variationally
 or by using quadratures is not so pronounced as $N$ increases. At the same time,
 for the Ar-Ar system the second order when calculated using quadratures fails
 to converge even for values of $N$ of the order of 400. Perhaps this is
 the best example in which the application of the KVP to correct the
 first order estimate is clearly necessary.
 In general we
 have observed that the scattering calculations need a bigger
 basis set to obtain converged results. This follow from the fact that
 the Ne-Ne and the Ar-Ar interactions predict
 an excited state close to zero energy.

 Finally we would like to mention the possibility of extending the present
 study to the three-body problem. In this case the dimension of the 
 matrices needed for the description of the processes are much bigger. 
 So the capability of producing convergence patterns in terms of $N$ could be limited.
 Hence variational estimates such as those presented here could help to 
 improve the DVR predictions. A study of this subject is at present in progress.

\newpage 

\begin{table}[!h]
\caption{Spectrum of the He, Ne and Ar dimers calculated with the
LM2M2, HFD-B and HFD-C potentials, respectively.}
\begin{center}\begin{tabular}{c|rrr} \hline \hline
      &  He(mK) & Ne(K)   & Ar(K)    \\ \hline
 $E_0$& -1.3020 & -24.4422 & -121.8571 \\
 $E_1$&         &  -4.5281 & -84.9229 \\
 $E_2$&         &  -0.0327 & -55.4794 \\
 $E_3$&         &         & -33.1574 \\
 $E_4$&         &         & -17.4289 \\
 $E_5$&         &         &  -7.5579 \\
 $E_6$&         &         &  -2.3509 \\
 $E_7$&         &         &  -0.3409 \\
 $E_8$&         &         & -1.9 10$^{-4}$ \\
\hline \hline
\end{tabular}
\end{center}\end{table}

\begin{table}[!h]
\caption{The He$_2$ binding energy, potential energy and mean square root radius,
 calculated with the VBR, DVR and $\bra \rm {DVR} \ket$ methods, 
 as a function of the size $N$ of the
 basis set employed. Energies are in mK, the radii in a.u.}
\begin{center}\begin{tabular}{c|rrr|rrr|rrr|} \hline \hline
 & \multicolumn{3}{c|}{VBR}& \multicolumn{3}{c|}{DVR} &  \multicolumn{3}{c|}{$\bra$DVR$\ket$} \\ \hline
 N & \multicolumn{1}{c}{E} & \multicolumn{1}{c}{V} & \multicolumn{1}{c|}{$\sqrt{\bra r^2 \ket}$} & \multicolumn{1}{c}{E} & \multicolumn{1}{c}{V} & \multicolumn{1}{c|}{$\sqrt{\bra r^2 \ket}$} & \multicolumn{1}{c}{E} & \multicolumn{1}{c}{V} & \multicolumn{1}{c|}{$\sqrt{\bra r^2 \ket}$} \\
    50 &      -0.0588 &     -141.786 &        67.34 &      -0.0742 &     -140.622 &        67.65 &       0.1002 &     -140.448 &        67.65 \\ 
   100 &      -1.2261 &     -108.303 &       106.66 &      -1.2337 &     -108.480 &       106.64 &      -1.2177 &     -108.464 &       106.64 \\ 
   150 &      -1.2943 &     -102.049 &       125.33 &      -1.2938 &     -102.034 &       125.34 &      -1.2942 &     -102.035 &       125.34 \\ 
   200 &      -1.3012 &     -100.904 &       131.86 &      -1.3026 &     -100.958 &       131.80 &      -1.3012 &     -100.957 &       131.80 \\ 
   250 &      -1.3019 &     -100.718 &       133.62 &      -1.3025 &     -100.742 &       133.59 &      -1.3019 &     -100.742 &       133.59 \\ 
   300 &      -1.3020 &     -100.690 &       134.01 &      -1.3009 &     -100.648 &       134.06 &      -1.3020 &     -100.649 &       134.06 \\ 
   350 &      -1.3020 &     -100.686 &       134.09 &      -1.3027 &     -100.714 &       134.05 &      -1.3020 &     -100.713 &       134.05 \\ 
   400 &      -1.3020 &     -100.686 &       134.10 &      -1.3014  &    -100.662 &       134.13  &     -1.3020 &     -100.662 &       134.13 \\
   450 &      -1.3020 &     -100.686 &       134.10 &      -1.3015  &    -100.665 &       134.13  &     -1.3020 &     -100.665 &       134.13 \\
   500 &      -1.3020 &     -100.686 &       134.11 &      -1.3019  &    -100.682 &       134.11  &     -1.3020 &     -100.682 &       134.11 \\
\hline \hline
\end{tabular}
\end{center}\end{table}

\newpage

\begin{table}[!h]
\caption{Convergence of the second order estimates of the phase shift $\delta$ 
(in radians) at different energies for the He-He, Ne-Ne and Ar-Ar systems, 
as a function of the dimension $N$ of the basis. The calculations using
the VBR, the DVR$_q$ and the $\bra {\rm DVR}\ket$ methods are compared.}
\begin{center}\begin{tabular}{c|rrr} \hline \hline
 N & VBR & DVR & $\bra {\rm DVR} \ket $ \\ \hline
 \multicolumn{4}{c}{He-He (E=50 mK)} \\ \hline
   20 &       1.2563 &      -1.3694 &      -1.5563 \\ 
   40 &       1.4685 &       1.4476 &       1.4627 \\ 
   60 &       1.4860 &       1.4892 &       1.4741 \\ 
   80 &       1.4767 &       1.4635 &       1.4784 \\ 
  100 &       1.4883 &       1.4874 &       1.4877 \\ 
  120 &       1.4883 &       1.4884 &       1.4882 \\ 
  140 &       1.4883 &       1.4879 &       1.4883 \\ 
  160 &       1.4883 &       1.4884 &       1.4883 \\ 
  180 &       1.4883 &       1.4882 &       1.4883 \\ 
  200 &       1.4883 &       1.4884 &       1.4883 \\  \hline
 \multicolumn{4}{c}{Ne-Ne (E=50 mK)} \\ \hline
   20 &       1.0573 &       1.5592 &      -1.5708 \\ 
   40 &       1.3272 &       1.3450 &      -0.0693 \\ 
   60 &       1.3309 &       1.3456 &       1.3209 \\ 
   80 &       1.3310 &       1.3311 &       1.3309 \\ 
  100 &       1.3310 &       1.3314 &       1.3310 \\ 
  120 &       1.3310 &       1.3310 &       1.3310 \\ 
  140 &       1.3310 &       1.3310 &       1.3310 \\ \hline 
 \multicolumn{4}{c}{Ar-Ar (E=2 mK)} \\ \hline
   10 &       1.2477 &      -0.1641 &       1.4324 \\ 
   50 &      -0.4193 &      -0.6421 &      -1.5435 \\ 
  100 &       0.9310 &       1.1382 &      -1.4981 \\ 
  150 &       1.0099 &       1.0077 &       0.9992 \\ 
  200 &       1.0109 &       1.0099 &       1.0108 \\ 
  250 &       1.0110 &       1.0107 &       1.0111 \\ 
  300 &       1.0110 &       1.0108 &       1.0111 \\ 
  350 &       1.0110 &       1.0108 &       1.0111 \\ 
\hline \hline
\end{tabular}
\end{center}\end{table}

\newpage

\noindent Figure Captions

\noindent Figure 1. 

 Ground state energy of He$_2$, calculated using the VBR (solid line), DVR(dashed
 line) and $\bra {\rm DVR} \ket$ (dotted line) methods, as a function of the 
 dimension $N$ of the basis. 

\noindent Figure 2. 
 (a) ground state energy of Ne$_2$, (b) first excited state energy of Ne$_2$,
 (c) ground state energy of Ar$_2$ and (d) eighth level of Ar$_2$,
 calculated using VBR (solid line), DVR (dashed line) and $\bra {\rm DVR} \ket$
 (dotted line), as a function of the dimension $N$ of the basis.
 For clarity the results have been drawn with 
 a solid line (VBR), a dashed line (DVR) and a dotted line
 ($\bra {\rm DVR} \ket$). 

\noindent Figure 3. 
 Convergence patterns for the energy of the ground state of Ne$_2$ (a), 
 the first excited vibrational state of  Ne$_2$ (b), and the eighth level
 of Ar$_2$ (c), for different choices of the size $n'$ of the restricted 
 variational problem of eq. (\ref{varham}). For Ne$_2$ the lines
 correspond to $n'=4$ (solid), $n'=8$ (dashed) and $n'=12$ (dotted-dashed).
 For Ar$_2$ the lines
 correspond to $n'=12$ (solid), $n'=20$ (dashed) and $n'=30$ (dotted-dashed).
 The VBR results are shown as a dotted line for reference.

\noindent Figure 4. 
Second order estimate for the He-He scattering length $a$ calculated with 
the VBR (solid line),
$\bra {\rm DVR}\ket$ (dashed line) and DVR$_q$ (dotted line) as
a function of $N$ (left panel). 
First order estimate for $a$, as obtained with the VBR method (solid line), 
and DVR (dotted line) as a function of $N$ (right panel).

\noindent Figure 5. 
Second order estimate for the Ne-Ne scattering length $a$ calculated with 
the VBR (solid line), $\bra {\rm DVR}\ket$ (dashed line) and DVR$_q$ 
(dotted line) as a function of $N$ (left panel). 
Second order estimate for the Ar-Ar scattering length $a$ calculated with 
the VBR (solid line), $\bra {\rm DVR}\ket$ (dashed line) and DVR$_q$ 
(dotted line) as a function of $N$ (right panel). 
 
\noindent Figure 6.
Zero energy DVR wave functions for He$_2$ (upper panel), Ne$_2$ (middle panel)
and Ar$_2$ (lower panel) calculated  using $N=300$, $N=200$ and $N=400$ respectively.

\noindent Figure 7. 
The r.h.s. of eq. (\ref{efr}) (solid line) compared to $q/{\rm tan}\delta$ 
(filled circles) for He-He (upper panel), Ne-Ne (middle panel) and Ar-Ar
(lower panel).

\newpage

\begin{figure}
\includegraphics[scale=0.7]{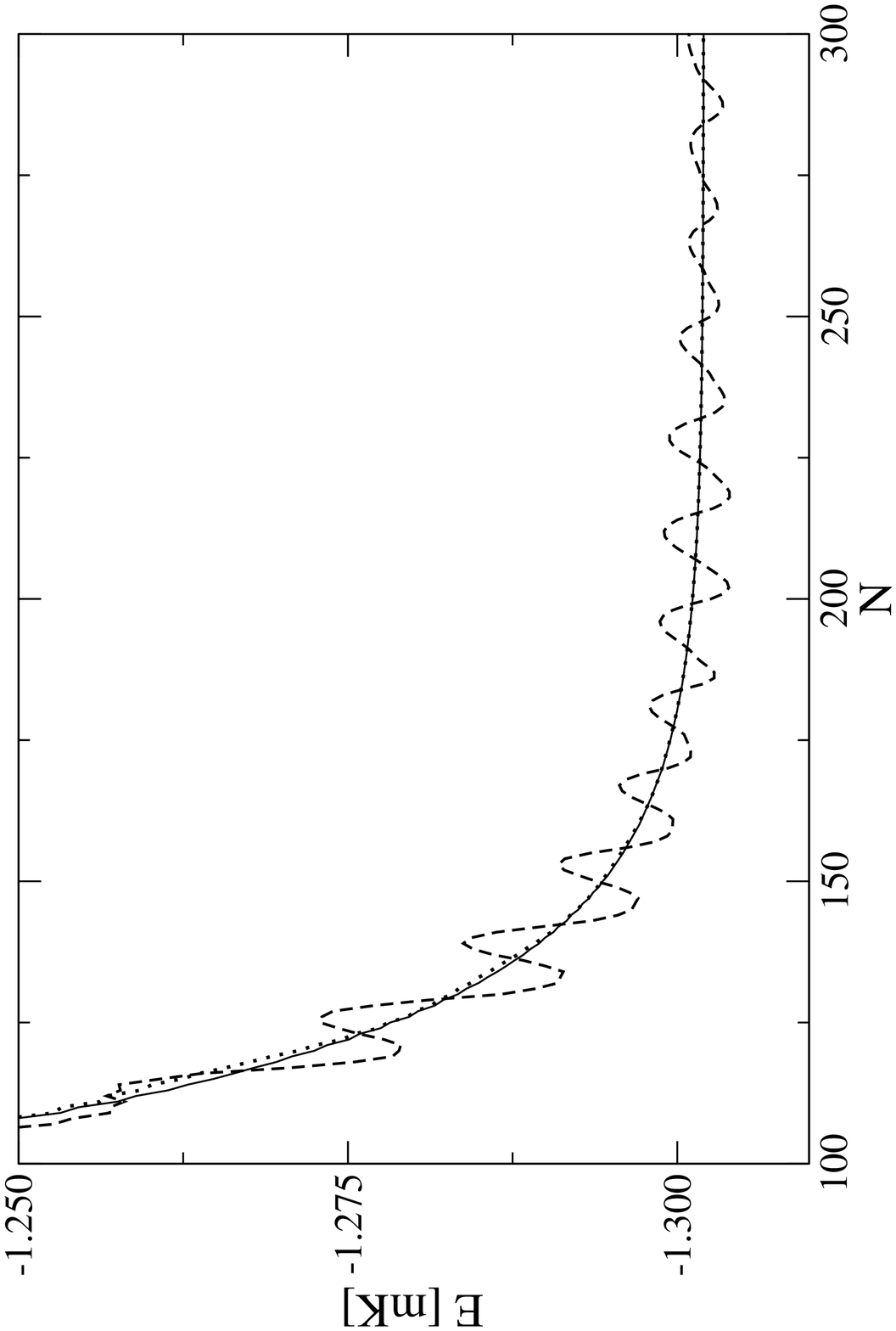}
\end{figure}

\newpage

\begin{figure}
\includegraphics[scale=0.8]{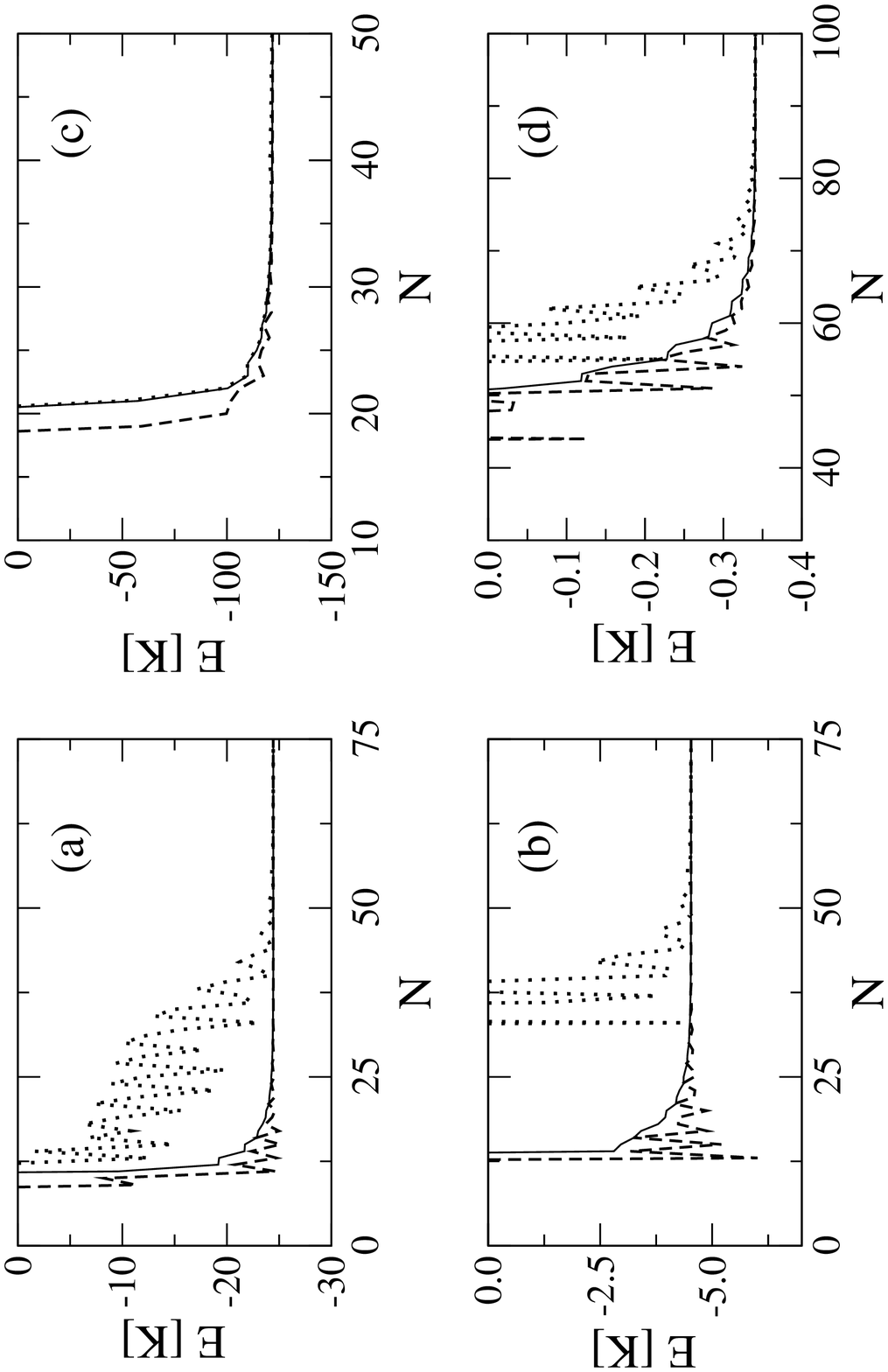}
\end{figure}

\newpage

\begin{figure}
\includegraphics[scale=1,angle=-90]{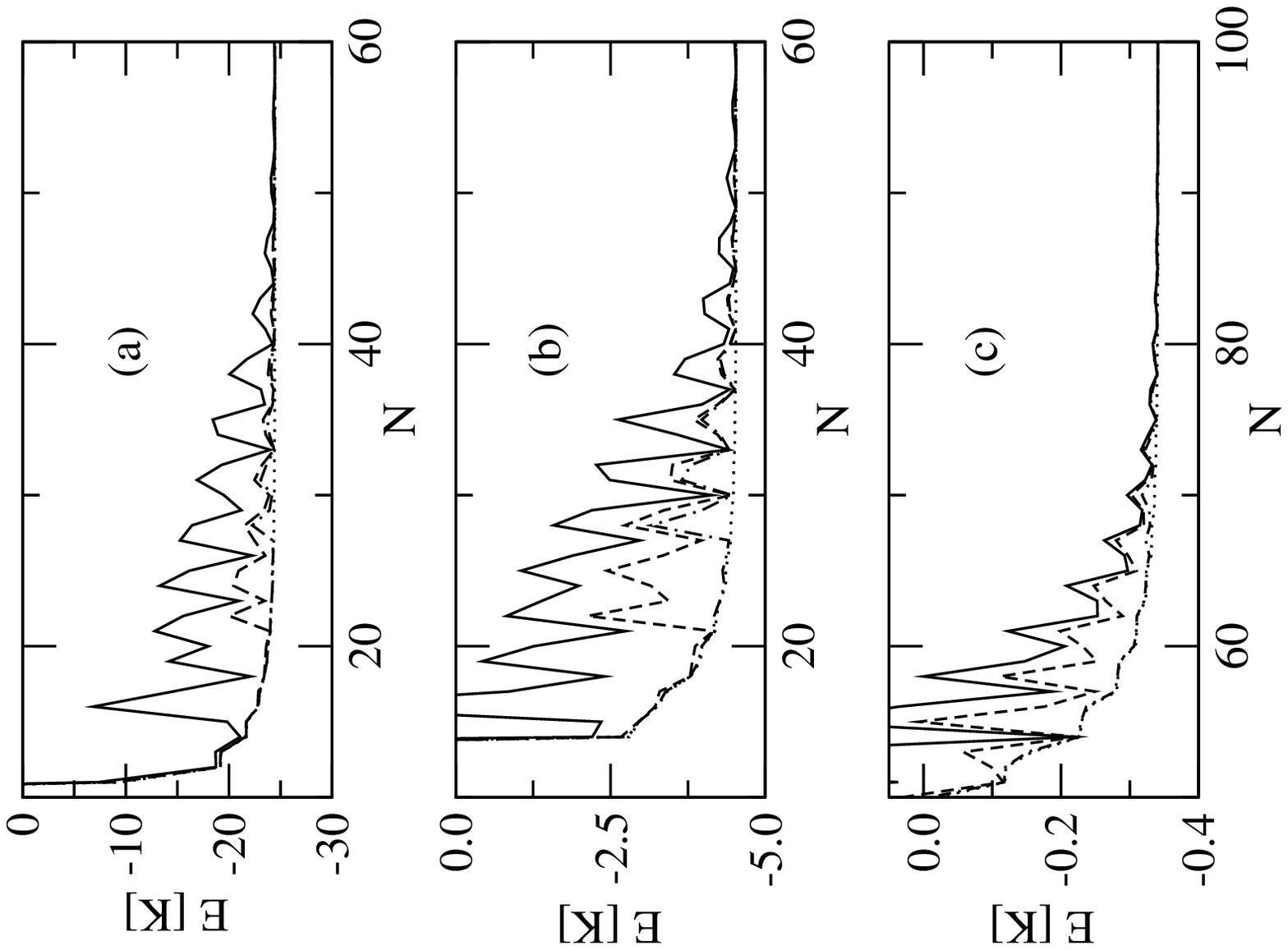}
\end{figure}

\newpage

\begin{figure}
\includegraphics[scale=0.8]{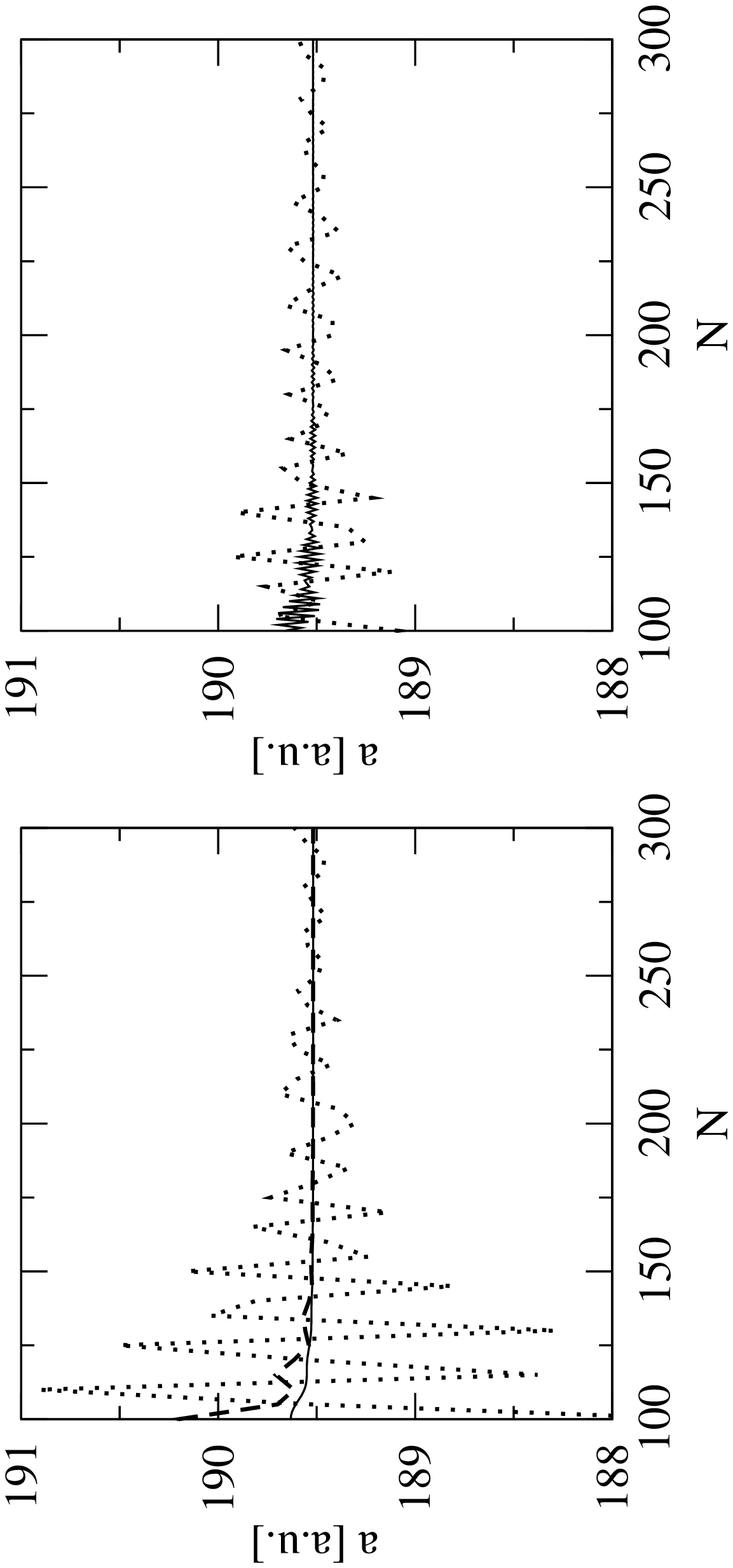}
\end{figure}

\newpage

\begin{figure}
\includegraphics[scale=0.8]{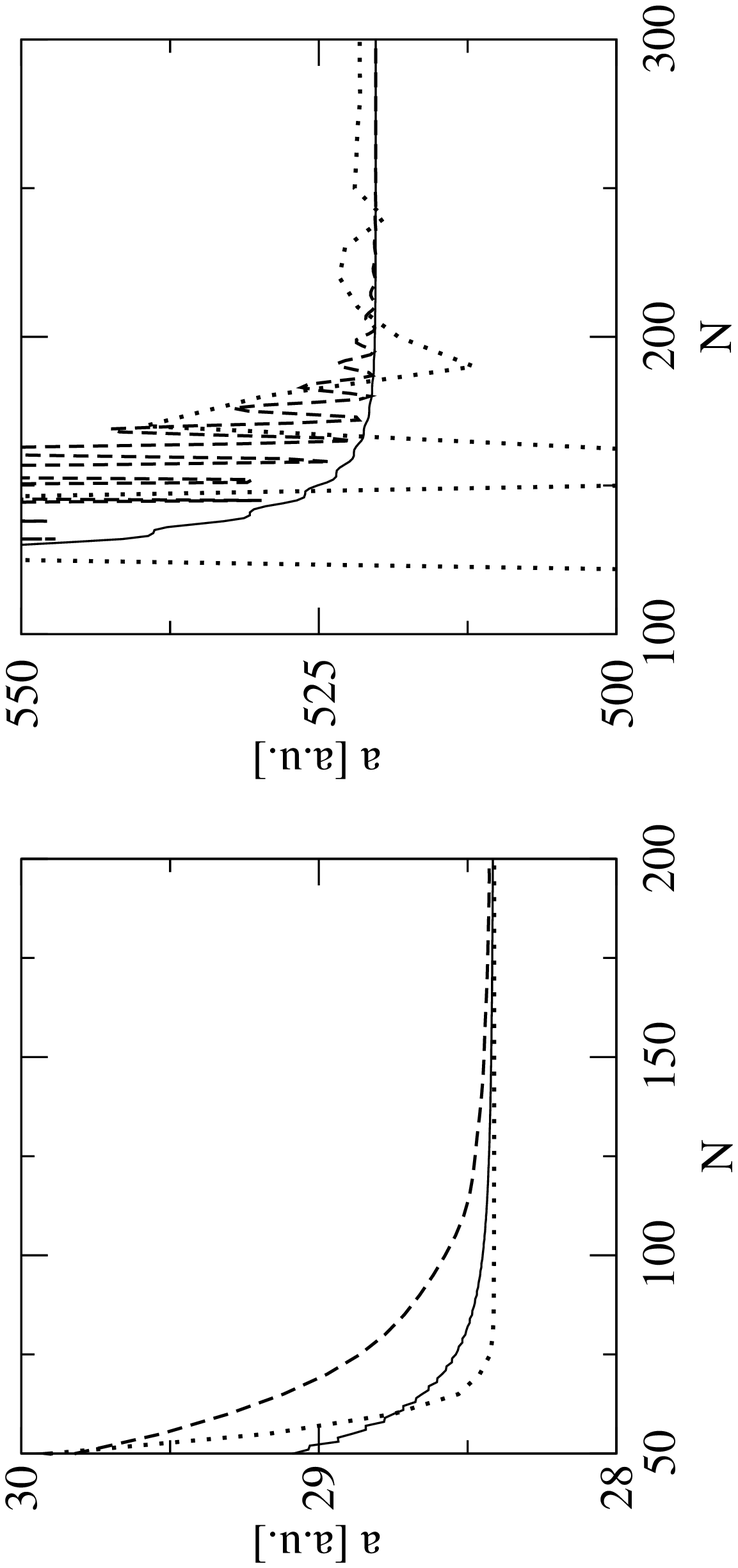}
\end{figure}

\newpage

\begin{figure}
\includegraphics[scale=0.8]{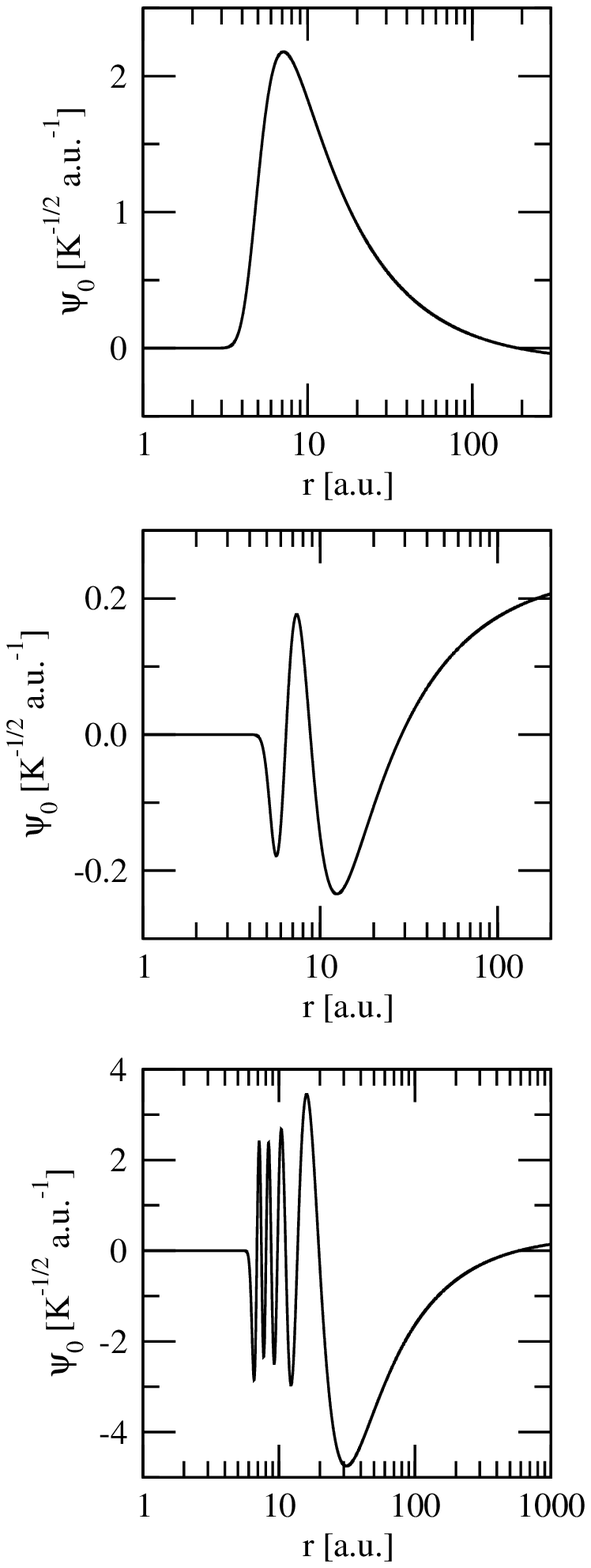}
\end{figure}

\newpage

\begin{figure}
\includegraphics[scale=0.8]{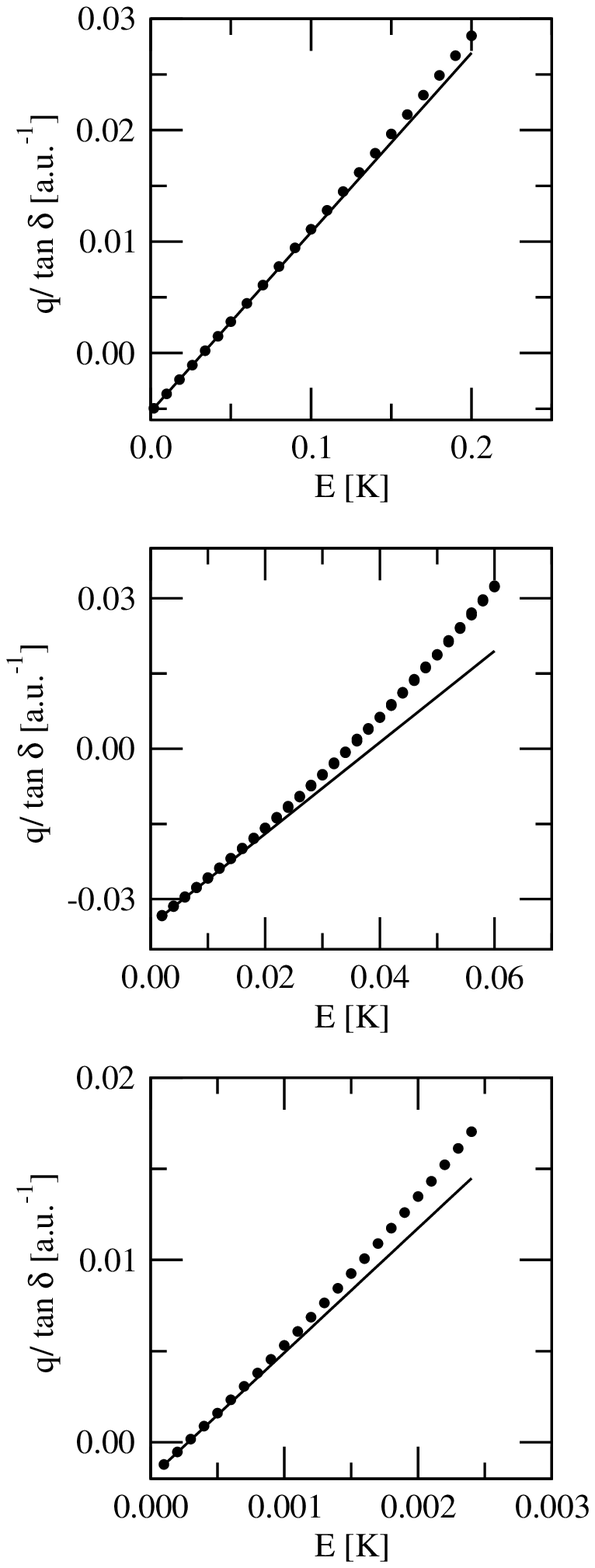}
\end{figure}

\end{document}